\documentclass[twocolumn]{jpsj3}

\usepackage{graphicx,color}
\usepackage{amsmath}
\usepackage{amssymb}
\usepackage{bm}
\usepackage{ulem}

\usepackage{txfonts}

\def\Hc2{B_\mathrm{c2}}
\def\214{\mathrm{Sr_2RuO_4}}
\def\327{\mathrm{Sr_3Ru_2O_7}}
\def\Tc{T_\mathrm{c}}
\def\muup{\mu}

\begin{document}

\title{High-Resolution Magnetostriction Measurements \\of the Pauli-Limited Superconductor Sr$_2$RuO$_4$}

\author{
	Shunichiro \textsc{Kittaka}$^{1,2}$\thanks{kittaka@g.ecc.u-tokyo.ac.jp}, 
	Yohei \textsc{Kono}$^{2}$, 
	Toshiro \textsc{Sakakibara}$^{3}$, 
	Naoki \textsc{Kikugawa}$^{4}$, 
	Shinya \textsc{Uji}$^{4}$, \\
	Dmitry A. \textsc{Sokolov}$^{5}$, and 
	Kazushige \textsc{Machida}$^{6}$ 
}

\inst{$^{1}$Department of Basic Science, The University of Tokyo, Meguro, Tokyo 153-8902, Japan\\
      $^{2}$Department of Physics, Faculty of Science and Engineering, Chuo University, Kasuga, Bunkyo-ku, Tokyo 112-8551, Japan\\
      $^{3}$Institute for Solid State Physics, The University of Tokyo, Kashiwa, Chiba 277-8581, Japan\\
      $^{4}$National Institute for Materials Science, 3-13 Sakura, Tsukuba, Ibaraki 305-0003, Japan\\
      $^{5}$Max Planck Institute for Chemical Physics of Solids, Nothnitzer Str. 40, 01187 Dresden, Germany\\
      $^{6}$Department of Physics, Ritsumeikan University, Kusatsu, Shiga 525-8577, Japan
}
\date{\today}

\abst{
We performed high-resolution magnetostriction measurements on the Pauli-limited superconductor $\214$ using high-quality single crystals. 
A first-order superconducting transition, accompanied by pronounced hysteresis, was observed under in-plane magnetic fields, 
where the relative length change of the sample, $\Delta L/L$, was on the order of $10^{-8}$.
To ensure the reliability of the measurements, particular attention was paid to minimizing the influence of magnetic torque,
which can significantly affect data under in-plane field configurations, via field-angle-resolved magnetostriction.
Within the hysteresis regime, slightly below the Pauli-limited upper critical field,
a hump-like anomaly in the magnetostriction coefficient was identified.
Furthermore, a characteristic double-peak structure in the field-angle derivative of the magnetostriction provides additional support for this anomaly.
Although these findings may reflect a lattice response associated with the emergence of the Fulde-Ferrell-Larkin-Ovchinnikov (FFLO) phase in $\214$,
the possibility of a broadened first-order transition cannot be excluded.
Notably, this magnetostriction anomaly qualitatively deviates from the FFLO phase boundary suggested by previous NMR measurements,
highlighting the necessity for further experimental and theoretical investigations to elucidate the nature of the FFLO state in this material.
}

\maketitle

\section{Introduction}
Theoretical proposals for a spatially modulated superconducting state induced by the Zeeman effect, 
known as the Fulde-Ferrell-Larkin-Ovchinnikov (FFLO) state \cite{Fulde1964PR,Larkin1964ZETF}, 
have inspired extensive experimental efforts to observe this exotic phase.
Despite these efforts, experimental evidence for the FFLO state has been reported only in a limited number of candidate materials, 
such as the organic superconductor $\kappa$-(BEDT-TTF)$_2$Cu(NCS)$_2$ \cite{Mayaffre2014NatPhys,Agosta2017PRL}, the heavy-fermion superconductor CeCoIn$_5$ \cite{Bianchi2003PRL}, and iron-based superconductors.\cite{Cho2017PRL,Kasahara2020PRL}
This limitation is likely due to the stringent conditions required for the emergence of the FFLO state \cite{Matsuda2007JPSJ}:
(i) a highly clean system, (ii) low dimensionality that enhances Fermi-surface nesting, and 
(iii) a strong Pauli-paramagnetic effect that surpasses the orbital depairing.
In superconductors with strong Pauli limiting, 
the nature of the superconducting transition under magnetic field $B$ 
changes from second order to first order upon cooling, typically below $0.56\Tc$ \cite{Matsuda2007JPSJ}.

These criteria are fulfilled in the layered perovskite superconductor $\214$ \cite{Mackenzie2003RMP,Maeno2012JPSJ,Maeno2024JPSJ,Maeno2024NP}.
First, the material exhibits exceptionally high purity, with a mean free path of approximately 1 $\muup$m \cite{Mackenzie1998PRL}, which far exceeds the in-plane coherence length $\xi_a$ of 660~$\AA$. 
Second, its Fermi surface topology \cite{Bergemann2000PRL,Damascelli2000PRL}, along with the intrinsic anisotropy in the coherence length ($\xi_{a}/\xi_c \sim 60$),\cite{Rastovski2013PRL,Kittaka2014PRB,Nakai2015PRB}
highlights its pronounced two-dimensionality, an essential feature for Fermi-surface nesting. 
Third, when a magnetic field is applied nearly parallel to the $ab$ plane, 
the upper critical field $\Hc2$ is significantly limited at low temperatures, and
a first-order superconducting transition occurs below approximately 0.8~K ($\sim 0.53\Tc$). 
This transition has been confirmed through various thermodynamic measurements, including specific heat \cite{Yonezawa2014JPSJ}, magnetocaloric effect \cite{Yonezawa2013PRL}, magnetization \cite{Kittaka2014PRB}, and magnetic torque \cite{Kittaka2014PRB,Kikugawa2016PRB}.
Indeed, thermodynamic estimation of the Pauli-limiting field, $B_{\rm P}=B_{\rm c}/\sqrt{\chi_{\rm n}-\chi_{\rm sc}} \sim 1.4$~T, is in good agreement with the observed limitation of $\Hc2$ for $B \parallel ab$. 
This estimation adopts a thermodynamic critical field of $B_{\rm c} = 0.02$~T, a spin susceptibility in the normal (superconducting) state of $\chi_{\rm n} = 0.9 \times 10^{-3}$ ($\chi_{\rm sc}=0$)~emu/mol, and a molar volume of $V_{\rm m}=57.5$ cm$^3$/mol.
Such consistency strongly suggests that the Pauli-paramagnetic effect is crucial for $\214$.

$\214$ has long been regarded as a promising candidate for a spin-triplet superconductor with a chiral $p$-wave order parameter \cite{Mackenzie2003RMP,Maeno2012JPSJ}
primarily based on NMR Knight shift and spin-polarized neutron scattering experiments that indicated 
an invariant spin susceptibility across $\Tc$ for all magnetic-field directions \cite{Ishida1998Nature,Ishida2008JPCS,Duffy2000PRL}.
However, this interpretation has become increasingly controversial following the discovery of a first-order transition at $\Hc2$,\cite{Yonezawa2013PRL} accompanied by a magnetization jump \cite{Kittaka2014PRB}. 
More critically, in 2019, Pustogow {\it et al}. reported a significant reduction in the NMR Knight shift in the superconducting state under low radio-frequency pulse powers \cite{Pustogow2019Nature}.
This result was subsequently confirmed by Ishida {\it et al}.\cite{Ishida2019} and further supported by polarized neutron scattering experiments.\cite{Petsch2020PRL} 
A similar $B$--$T$ phase diagram, featuring both a first-order phase transition and Knight-shift reduction, has also been observed 
in uniaxially strained samples with an enhanced $\Tc$ of approximately 3.4~K \cite{Pustogow2019Nature,Steppke2017Science}. 
These findings strongly suggest that the first-order phase transition for $B \parallel ab$ originates from the Pauli-paramagnetic effect, 
in stark contrast to the long-standing chiral $p$-wave scenario. 
Consequently, a paradigm shift is underway in the understanding of the superconducting order parameter in $\214$.\cite{Maeno2024JPSJ,Maeno2024NP}

Thus, an intriguing question arises as to whether the FFLO state is realized in $\214$.
A fourfold oscillation in the specific heat under an in-plane rotating magnetic field at low temperatures was found to be abruptly suppressed around 1.3~T,
followed by its drastic development above 1.4~T with an opposite sign,\cite{Kittaka2018JPSJ}
which may reflect the field-angle anisotropy associated with the FFLO state.
Recently, NMR measurements have provided compelling evidence for the emergence of the FFLO phase in $\214$.\cite{Kinjo2022Science}
Nevertheless, clear thermodynamic signatures of such a double transition remain elusive, 
despite long-standing discussions linking its possible existence to chiral triplet pairing~\cite{Mao2000PRL,Deguchi2002JPSJ,Yaguchi2002PRB},
particularly the splitting of degenerate order parameters in the chiral $p_x \pm ip_y$ state.\cite{Kaur2005PRB,Udagawa2005JPSJ,Mineev2008PRB}

\section{Methods}

In this study, we performed high-resolution field-angle-resolved magnetostriction and thermal-expansion measurements on $\214$ as a powerful experimental approach.
This technique offers a significant advantage for probing first-order phase transitions, as recently demonstrated in CeCoIn$_5$.\cite{Kittaka2023PRB}
We utilized a home-built capacitively-detected dilatometer, which achieves a resolution significantly better than 0.01~$\AA$. 
The relative change in sample length along the $i$ direction, $\Delta L_i(T,B)=L_i(T,B)-L_i(T_0,B_0)$, was determined 
from the change in capacitance, $\Delta C=C(T,B)-C(T_0,B_0)$, between the movable and fixed electrodes using the relation
$\Delta L_i = \varepsilon_0 A\Delta C/[C(T,B)C(T_0,B_0)]$,
where $A=25\pi$~mm$^2$ is the electrode area and $\varepsilon_0$ is the vacuum permittivity.
A typical capacitance value is approximately 13~pF, corresponding to a capacitor gap of about 0.05~mm.
The sample was gently sandwiched between the movable component and an adjustment screw,
the latter being firmly fixed to the outer frame with a locking nut. 
To prevent sample rotation due to magnetic torque, the sample was bonded to the adjustment screw using GE varnish.
For comparison, the specific heat $c_p$ was measured using the quasi-adiabatic heat-pulse method in a dilution refrigerator.
In all measurements, the magnetic-field orientation was precisely controlled in three dimensions using a vector-magnet system \cite{Sakakibara2016RPP}.

High-quality single crystals used in this study were grown by the floating-zone method \cite{Mao2000MRB,Bobowski2019CM}.
We selected a large single crystal weighing 50~mg, 
with approximate dimensions of 3 mm $\times$ 3 mm in the $ab$ plane and 1.1 mm along the $c$ axis,
exhibiting an onset $\Tc$ of 1.525~K.
Both surfaces along the $ab$ plane were polished, while the $bc$ plane was left as-cut.
Magnetostriction and thermal-expansion measurements were performed along the $c$ and $a$ axes in a $^3$He refrigerator or a dilution refrigerator, 
where $L_c \sim 1.1$~mm and $L_a \sim 2.7$~mm.
To obtain a large $\214$ sample with a high $\Tc$, a small amount of bilayer $\327$ inclusions was unavoidable.
These inclusions were identified via specific-heat measurements and polarized light optical microscopy  of the polished surface (see Supplemental Material~\cite{KittakaSM}). 
Since $c_p/T$ of $\327$ in the $\214$-$\327$ eutectic system is known to be temperature-independent below 2~K \cite{Kittaka2008PRB},
we subtract its contribution from the data presented below~\cite{KittakaSM}. 
The corrected specific heat is referred to as $c_{214}$.
Here, the phonon contribution, assuming a Debye temperature of 410~K, and the nuclear specific heat, given by $C_N(T,B)=(0.08+0.14B^2)/T^2$~$\muup$J / (mol K)~\cite{Kittaka2018JPSJ}, are subtracted 
along with the addenda heat capacity, which is always less than 5\% of the sample heat capacity.
Moreover, $\327$ inclusions are expected to have only a minor effect on the magnetostriction, 
as no anomaly has been reported for $\327$ in the low-field region below 2~T \cite{Perry2005JPSJ}.
To confirm the reproducibility of our results, we also measured $\Delta L_c$ for a relatively small, pure $\214$ sample (12~mg),
which was free of $\327$ inclusions~\cite{KittakaSM}.

\section{Results and Discussion}
\subsection{Specific heat}
\begin{figure}
\includegraphics[width=3.2in]{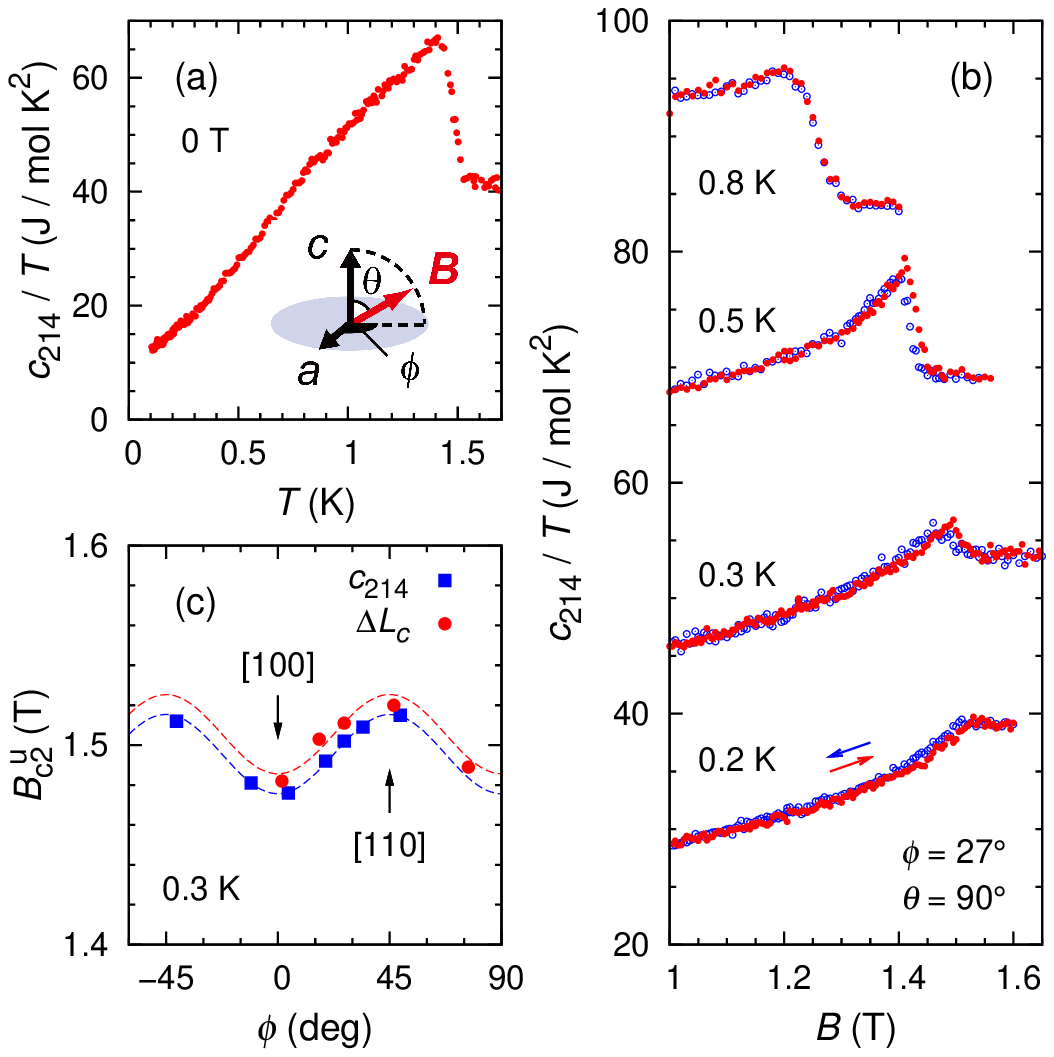} 
\caption{
(Color online) (a) Temperature dependence of the zero-field specific heat of the $\214$ component, $c_{\rm 214}$, divided by temperature. 
(b) Magnetic field dependence of $c_{\rm 214}/T$ at several temperatures, measured at $\phi=27^\circ$ and $\theta=90^\circ$.
Each dataset at the same temperature is vertically shifted by 10~mJ/(mol K$^2$) for clarity.
The blue open and red closed circles represent data obtained during decreasing and increasing field sweeps, respectively.
(c) Field-angle $\phi$ dependence of the upper critical field $\Hc2$ at $\theta=90^\circ$ (i.e., within the $ab$ plane) at 0.3~K, 
determined from specific-heat and magnetostriction $\Delta L_c$ measurements during increasing field sweeps.
The inset in (a) depicts the definition of the field angles, $\phi$ and $\theta$.
}
\label{SH}
\end{figure}

Figure \ref{SH}(a) shows the temperature dependence of the zero-field electronic specific-heat data, $c_{214}/T$, for the $\214$ component.
A clear specific-heat jump is observed at $\Tc=1.483$~K (midpoint), with an onset temperature of 1.525~K.
The width of the jump is less than 0.09~K, and no additional anomalies or kinks are present.
These results demonstrate the high quality and narrow $\Tc$ distribution of the present sample,
despite the relatively large sample size.

Figure \ref{SH}(b) presents the magnetic-field dependence of $c_{214}/T$ at several temperatures under an in-plane magnetic field ($\theta=90^\circ$) with $\phi=27^\circ$.
Here, as illustrated in the inset of Fig.~\ref{SH}(a), $\phi$ and $\theta$ denote the azimuthal and polar angles of the magnetic field, measured from the $[100]$ and $[001]$ axes, respectively.
These angles were defined based on the anisotropy of $\Hc2$ [Fig.~\ref{SH}(c)] and verified using Laue diffraction images.
Despite the relatively large sample size, a clear hysteresis is observed in $c_{214}(B)$ near $\Hc2$ at low temperatures, 
indicating the presence of a first-order superconducting transition.
This observation confirms that the $\214$ component in the present sample possesses sufficiently high quality.

As shown in Fig.~\ref{SH}(b), a specific-heat peak appears near $\Hc2$ in $c_{214}(B)$ at moderate temperatures.
This peak complicates the identification of possible internal phase transitions within the superconducting state, particularly at intermediate temperatures.
In addition, the specific-heat data exhibit some scatter, and the measurement sensitivity is not sufficiently high
to resolve subtle anomalies, such as those expected from the FFLO phase transition. 

\subsection{Thermal expansion}
\begin{figure}
\includegraphics[width=3.2in]{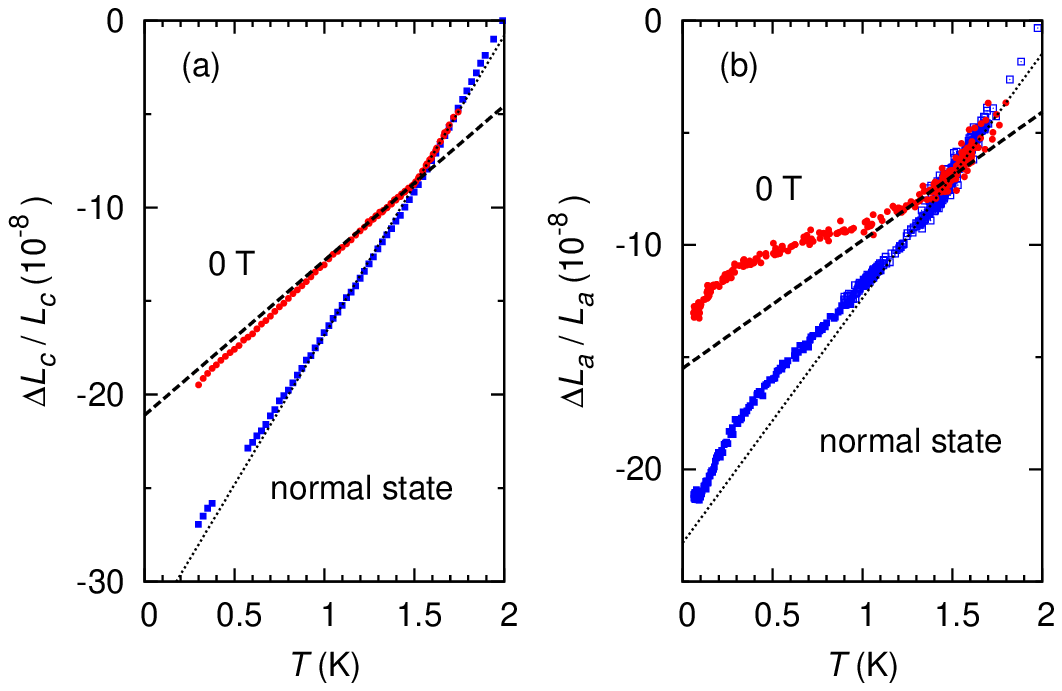} 
\caption{
(Color online) Temperature dependence of the relative length change, $\Delta L_i/L_i$, along the (a) $c$ and (b) $a$ axes.
The red circles represent zero-field data.
The blue squares in~(a) indicate normal-state data measured under an in-plane magnetic field of 2~T at $\phi=27^\circ$.
The open and closed squares in (b) correspond to normal-state data measured under in-plane fields of 1.45 and 1.8~T, respectively, at $\phi=90^\circ$.
The dashed (dotted) lines represent linear fits to the zero-field data just below (above) $\Tc$.
}
\label{T}
\end{figure}

To investigate possible internal superconducting phase transitions, we focus on 
the lattice response of the sample using a home-built, capacitively detected subpicometer dilatometer.
Figure~\ref{T} presents the thermal-expansion data.
A clear kink is observed at $\Tc=1.515$~K in both $L_c(T)$ and $L_a(T)$, consistent with the specific-heat results.
Here, the linear thermal-expansion coefficient along each axis is defined as $\alpha_i=(\partial L_i/\partial T)/L_i$. 
Across the superconducting transition, the coefficient exhibits a discontinuity $\Delta \alpha_i$.
As indicated by the dashed and dotted lines in Fig.~\ref{T}, we estimated 
the discontinuities in thermal expansion coefficients at $\Tc$ as 
$\Delta\alpha_c=-(7.5 \pm 0.5) \times 10^{-8}$~K$^{-1}$ and $\Delta\alpha_a=-(5.0 \pm 0.7)\times 10^{-8}$~K$^{-1}$.

For a second-order phase transition, the Ehrenfest relation $\partial \Tc/\partial \sigma_{ii}=-\Tc V_{\rm m}\Delta\alpha_i/\Delta c$ should be satisfied. 
Using the obtained values of $\Delta \alpha_i$ and the specific heat jump at $\Tc$, $\Delta c\sim 30$~mJ mol$^{-1}$ K$^{-2}$, in zero field,
we estimate $\partial\Tc/\partial\sigma_{xx} \sim 0.1$~K/GPa and $\partial\Tc/\partial\sigma_{zz} \sim 0.15$~K/GPa in the zero-pressure limit for $\214$.
Assuming hydrostatic pressure as a combination of uniaxial components, the pressure dependence of $\Tc$ is given by 
$d\Tc/dP \sim -(2\partial\Tc/\partial\sigma_{xx}+\partial\Tc/\partial\sigma_{zz}) \sim -0.3$~K/GPa,
which agrees well with the previous experimental observations ($\approx -0.2$~K/GPa).\cite{Shirakawa1997PRB,Forsythe2002PRL}
However, the estimated value of $\partial\Tc/\partial\sigma_{xx}$ exhibits a sign opposite to that observed experimentally,\cite{Barber2019PRB}
whereas the estimate for $\partial\Tc/\partial\sigma_{zz}$ is in good agreement with recent uniaxial strain measurements along the $c$ axis.~\cite{Jerzmbeck2022NatCom}
These contrasting results suggest that the strain dependence of $\Tc$ in $\214$ is more complex than expected from simple thermodynamic relations.\cite{Mattoni2025arXiv}

\subsection{Effect of magnetic torque on our magnetostriction measurements}
\begin{figure}
\includegraphics[width=3.2in]{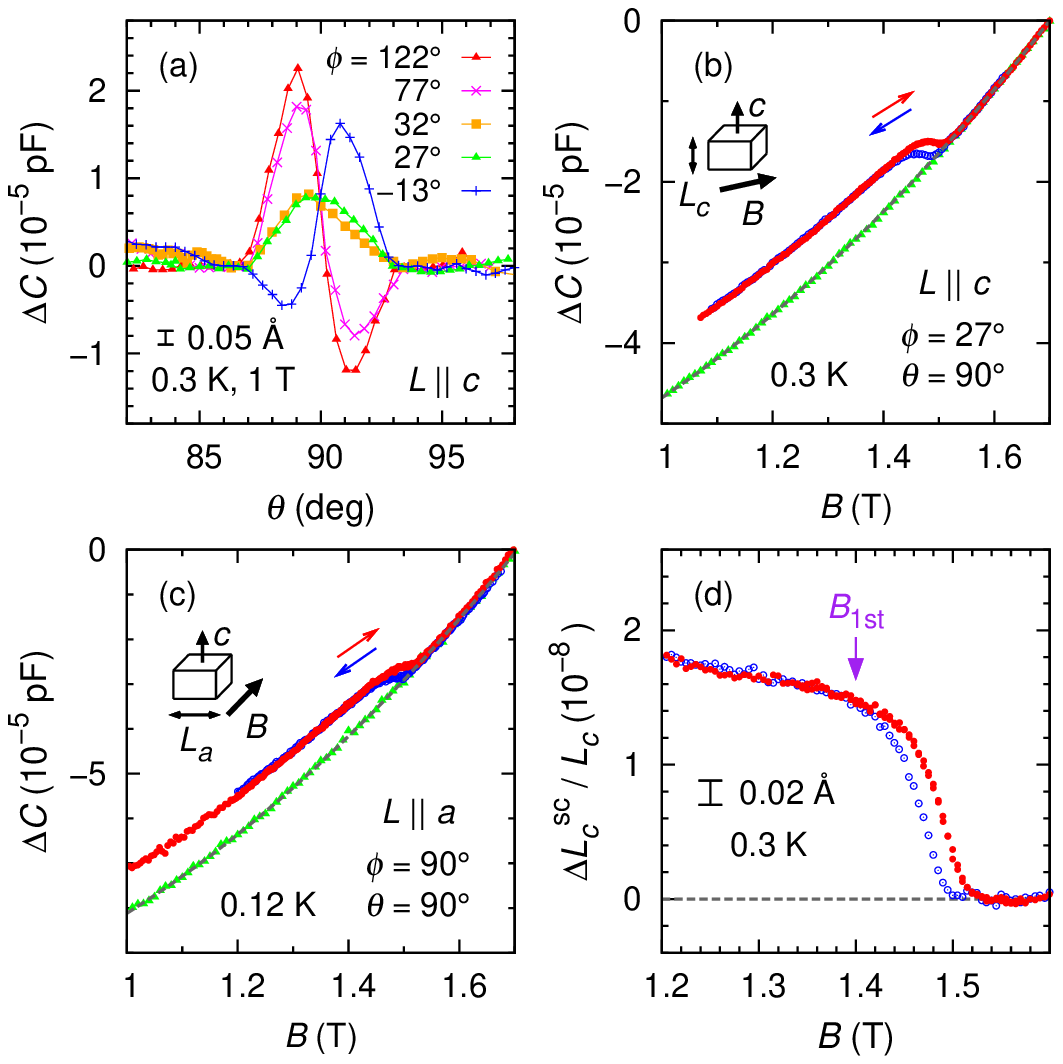} 
\caption{
(Color online) (a) Polar-angle $\theta$ dependence of the capacitance change relative to the normal-state value (taken at $\theta \sim 95^\circ$) for $L \parallel c$, measured at 0.3~K and 1~T for various azimuthal angles $\phi$.
(b), (c) Magnetic-field dependence of the capacitance change relative to the value at 1.7 T, measured at low temperatures for two different sample orientations:
(b) $L \parallel c$ at 0.3~K with $\phi = 27^\circ$ and $\theta = 90^\circ$, 
(c) $L \parallel a$ at 0.12~K with $\phi = 90^\circ$ and $\theta = 90^\circ$.
The blue open and red closed circles denote data obtained during decreasing and increasing field sweep.
The green triangles show normal-state response measured at $\theta = 85^\circ$ for each configuration, where $\Hc2$ is below 0.8~T. 
The dashed lines are fits to the normal-state data using a cubic polynomial function.
(d) Superconducting contribution to the normalized magnetostriction $\Delta L_c^{\rm sc}/L_c$ for $L \parallel c$, obtained by subtracting the background [dashed line in (b)]. 
The characteristic field $B_{\rm 1st}$, at which the hysteresis in $\Delta L_c^{\rm sc}$ vanishes, is indicated by an arrow.
}
\label{MagStr}
\end{figure}

Figure~\ref{MagStr}(a) shows the relative change in the dilatometer capacitance, $\Delta C$, for $L \parallel c$, as a function of the polar angle $\theta$ of the magnetic field ($B=1$~T) for various azimuthal angles $\phi$.
Although magnetostriction is expected to be symmetric with respect to $\theta=90^\circ$ due to the tetragonal crystal symmetry, 
$\Delta C$ exhibits clear asymmetry except near $\phi = 27^\circ$.
This asymmetric behavior is attributed to magnetic torque in the superconducting state, which reverses sign at $\theta = 90^\circ$~\cite{KittakaTorque},
and is likely due to the anisotropic mechanical response of our home-built dilatometer, where the movable electrode is suspended by crossed phosphor-bronze wires.
Indeed, a qualitatively similar $\theta$ dependence of magnetic torque has been reported for $\214$~\cite{Kittaka2014PRB}, 
originating from the large anisotropy in the coherence length ($\xi_a / \xi_c \sim 60$).~\cite{Kittaka2009JPCS-2}.
It is speculated that the torque induces a slight rotation of the sample, depending on the adhesion strength of the mounting paste. 
This rotation may shift the position of the movable electrode by approximately 0.1~$\AA$, 
which is comparable to the magnetostriction signal $\Delta L_c$ observed in the present sample.
Therefore, in the following analysis, we focus on magnetostriction data obtained at $\phi=27^\circ$, 
where the magnetic-torque effect is fortuitously minimized for $L \parallel c$ in this sample.

\subsection{Magnetostriction along the $c$ axis}
Figure~\ref{MagStr}(b) shows the magnetic-field dependence of $\Delta C(B)$ when $L \parallel c$, measured during increasing and decreasing field sweeps at 0.3~K with $\phi=27^\circ$ and $\theta=90^\circ$.
To estimate the non-superconducting background contribution, including the magnetostriction of the dilatometer itself, 
we also measured $\Delta C(B)$ in the normal state at $\theta=85^\circ$ (green triangles), where $\Hc2 < 0.8$~T~\cite{Kittaka2009PRB}.
By subtracting this background signal from the data at $\theta=90^\circ$, 
we isolate the superconducting contribution to the length change of the sample along the $i$ direction, defined as
$\Delta L_i^{\rm sc} \approx \varepsilon_0 A[\Delta C(\theta=90^\circ)-\Delta C(\theta=85^\circ)]/[C(T_0,H_0)]^2$.
The resulting normalized magnetostriction, $\Delta L_c^{\rm sc}/L_c$, is plotted in Fig.~\ref{MagStr}(d).
Remarkably, $\Delta L_c^{\rm sc}/L_c$ in $\214$ is on the order of $10^{-8}$, 
which is two orders of magnitude smaller than those observed in CeCoIn$_5$ \cite{Takeuchi2002JPCM,Correa2007PRL,Kittaka2023PRB} and CeCu$_2$Si$_2$,\cite{Weickert2018PRB} 
both of which exhibit values around $10^{-6}$.
These results highlight the importance of high-resolution magnetostriction measurements and careful consideration of parasitic magnetic-torque effects 
when investigating superconductivity in $\214$.

As shown in Figs.~\ref{MagStr}(b) and \ref{MagStr}(d), 
the onset of $\Hc2$ clearly differs between increasing and decreasing field sweeps.
This hysteresis indicates that the first-order superconducting transition at $\Hc2$ can be sensitively detected via magnetostriction.
As indicated by the arrow in Fig.~\ref{MagStr}(d), 
we can precisely define the characteristic field $B_{\rm 1st}$, at which the hysteresis loop in $\Delta L_i^{\rm sc}(B)$ closes within the superconducting phase.
Qualitatively similar behavior is observed in $\Delta L_c^{\rm sc}$ for another sample with $L_c=0.7$~mm, 
as shown in the Supplemental Material~\cite{KittakaSM}.

\begin{figure}
\includegraphics[width=3.2in]{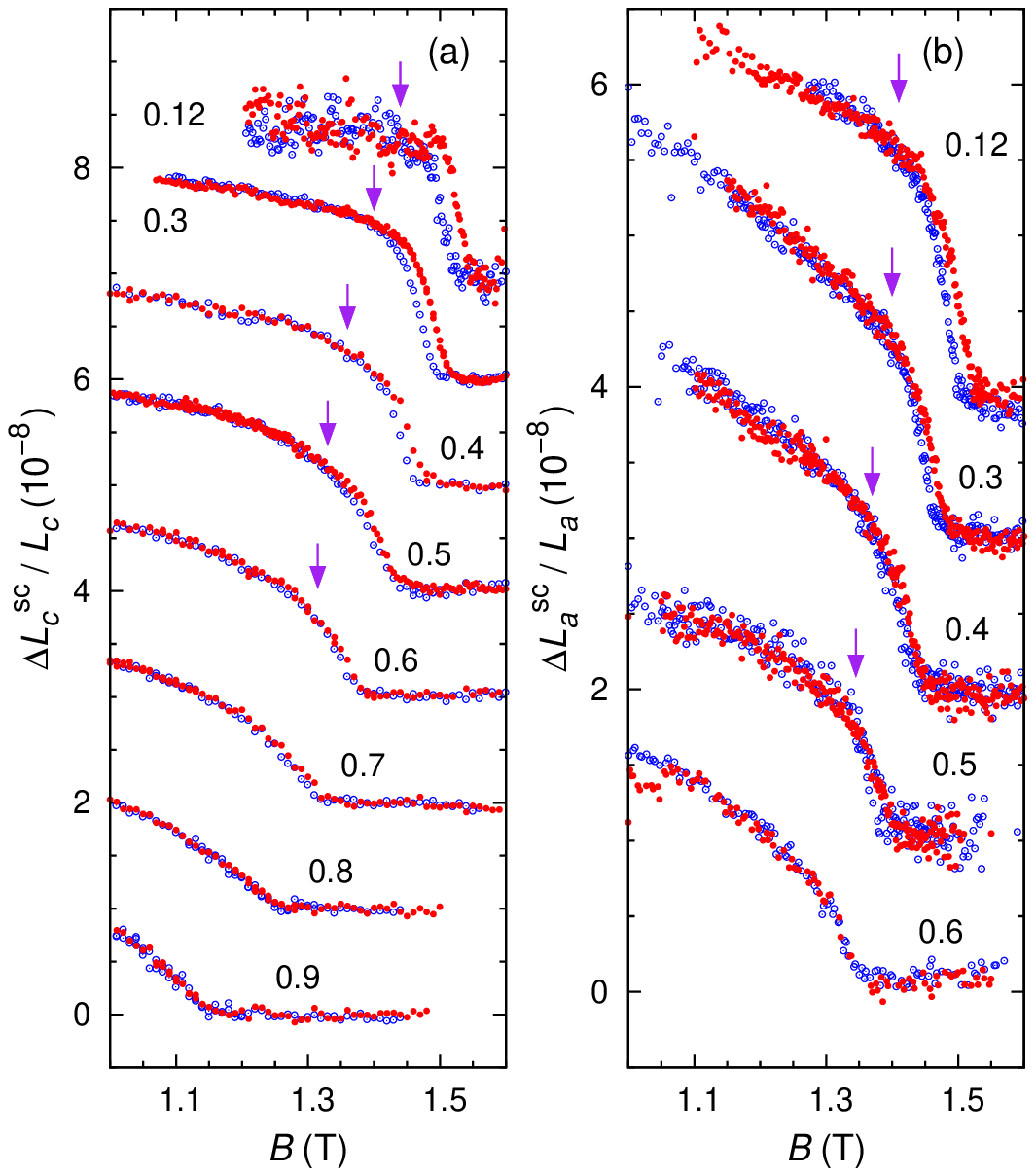} 
\caption{
(Color online) Magnetic-field dependence of the normalized magnetostriction $\Delta L_i^{\rm sc} / L_i$ at several temperatures for 
(a) $i=c$ ($\phi=27^\circ$ and $\theta=90^\circ$) and (b) $i=a$ ($\phi=90^\circ$ and $\theta=90^\circ$).
Each dataset is vertically shifted by $1 \times 10^{-8}$ for clarity.
The blue open and red closed circles represent data obtained during decreasing and increasing field sweeps, respectively.
The arrows indicate the characteristic field $B_{\rm 1st}$, at which the hysteresis loop closes.
The numbers labeling each dataset indicate the temperature in K. 
}
\label{Hdep}
\end{figure}

\begin{figure}[t]
\includegraphics[width=3.2in]{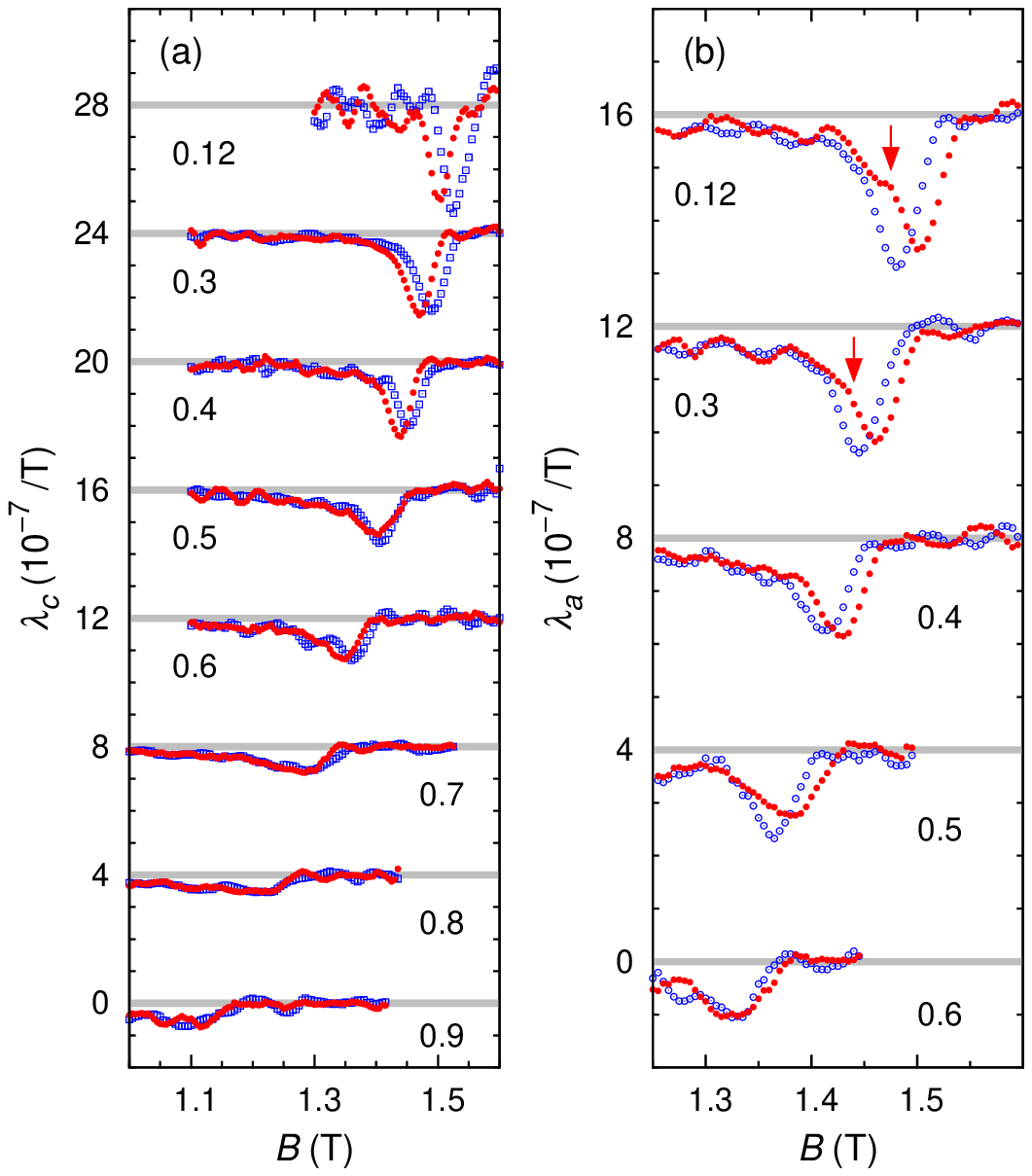} 
\caption{
(Color online) Magnetic-field dependence of the magnetostriction coefficient $\lambda_i=(\partial \Delta L_i^{\rm sc} / \partial B) /L_i$ at several temperatures for 
(a) $i=c$ ($\phi=27^\circ$ and $\theta=90^\circ$) and (b) $i=a$ ($\phi=90^\circ$ and $\theta=90^\circ$).
Each dataset is vertically shifted by $4 \times 10^{-7}$ T$^{-1}$ for clarity.
The blue open and red closed circles represent data obtained during decreasing and increasing field sweeps, respectively.
The arrows indicate a possible hump-like anomaly at $B_{\rm K}$.
The numbers labeling each dataset indicate the temperature in K. 
}
\label{Hdep2}
\end{figure}

Figure \ref{Hdep}(a) shows the magnetic-field dependence of $\Delta L_c^{\rm sc}/L_c$ at several temperatures.
The data at 0.12~K were measured in a dilution refrigerator, while those for $T\ge 0.3$~K were obtained using a $^3$He refrigerator.
The former dataset exhibits noticeable scatter, primarily due to heating of the sorption pump in the refrigerator.
As the temperature increases, $B_{\rm 1st}$ shifts toward lower values.
Above 0.6~K, it becomes difficult to determine $B_{\rm 1st}$ precisely because of the resolution limit of our dilatometer.
The magnetostriction coefficient, defined as $\lambda_i=(\partial \Delta L_i^{\rm sc}/\partial B)/L_i$, 
along the $c$ axis is plotted in Fig.~\ref{Hdep2}(a).

\subsection{Magnetostriction along the $a$ axis}
For $L \parallel a$, to avoid torque-related artifacts, we restrict our analysis to data taken at $\phi = 90^\circ$, where the magnetic-torque effect is minimal.
Figure~\ref{MagStr}(c) shows $\Delta C(B)$ for $L \parallel a$, measured during increasing and decreasing field sweeps at 0.12~K, with $\phi=90^\circ$ and $\theta=90^\circ$.
The superconducting contribution to the magnetostriction, $\Delta L_a^{\rm sc}/L_a$, and the corresponding magnetostriction coefficient $\lambda_a$ at several temperatures are shown in Figs.~\ref{Hdep}(b) and \ref{Hdep2}(b), respectively.
All data were obtained using a dilution refrigerator.

Theoretically, an abrupt enhancement of the zero-energy quasiparticle density of states, $N(E=0)$, has been predicted at the transition between the FFLO and Abrikosov states \cite{Suzuki2011JPSJ,Suzuki2019,Kittaka2023PRB}.
Such an anomaly should be observable through low-temperature thermodynamic quantities that reflect $N(E=0)$.
Indeed, previous specific-heat \cite{Yonezawa2014JPSJ} and entropy \cite{Yonezawa2013PRL} measurements at 0.3 K may have captured this anomaly, 
although the signature was not particularly pronounced.
Furthermore, recent NMR studies have revealed a characteristic double-horn spectrum at temperatures below 0.3 K and magnetic fields above approximately 1.2~T, 
indicating a spatial modulation of spin density intrinsic to the FFLO state. 
However, no corresponding anomaly was detected around 1.2~T in the present magnetostriction measurements.
Instead, a hump-like anomaly was detected in $\lambda_a(B)$ at $B_{\rm K}$, as indicated by the arrows in Fig.~\ref{Hdep2}(b), above $B_{\rm 1st}$.

\begin{figure}[t]
\includegraphics[width=3.2in]{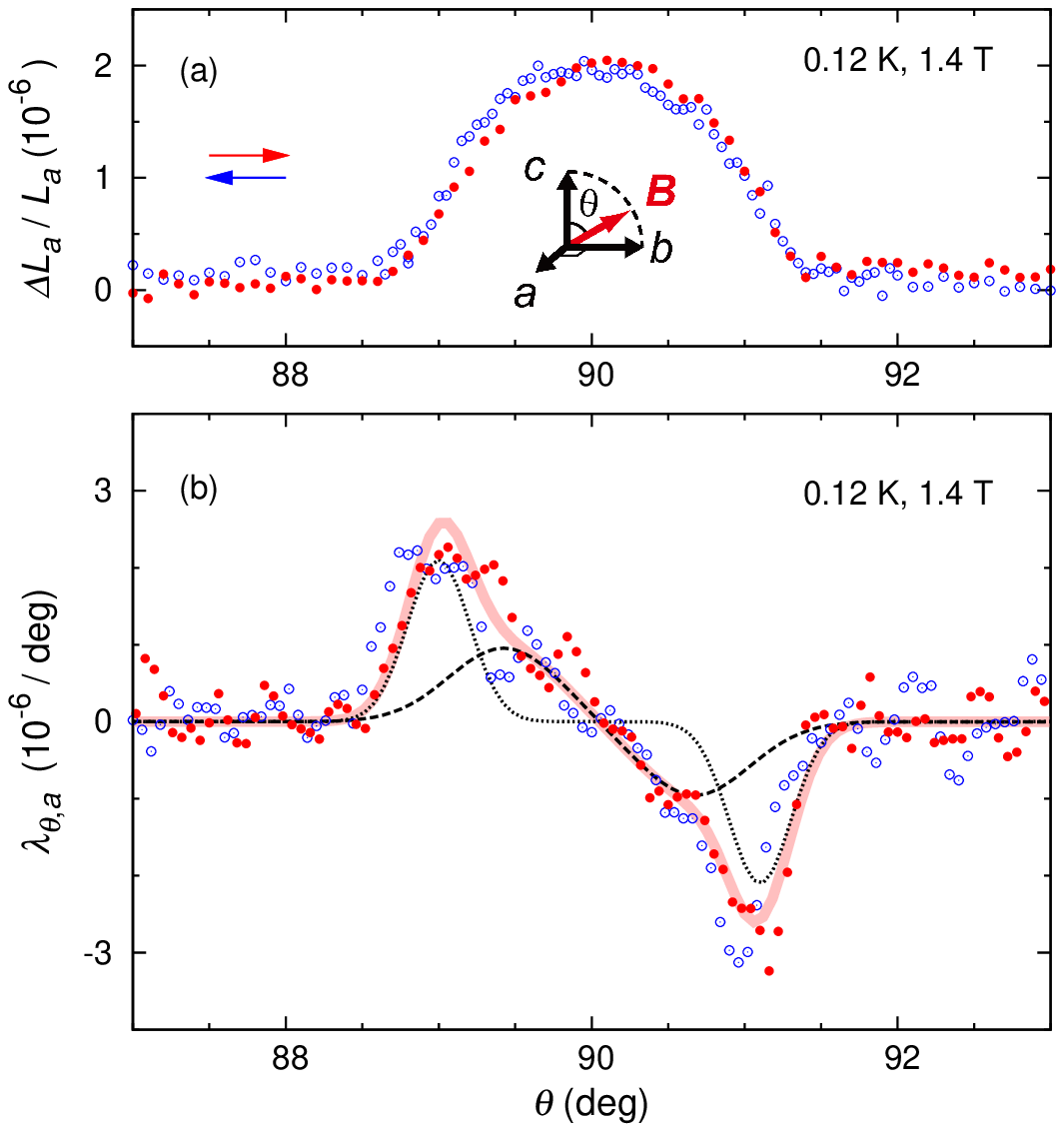} 
\caption{
(Color online) Field-angle $\theta$ dependence of (a) the magnetostriction $\Delta L_a/L_a$ and (b) its coefficient $\lambda_{\theta,a}=(\partial L_a/\partial \theta)/L_a$, measured at 0.12~K under a magnetic field of 1.4~T rotated within the $bc$ plane at $\phi=90^\circ$.
The blue open and red closed circles represent data obtained during decreasing and increasing $\theta$ sweeps, respectively.
The solid line in (b) represents a fit to the data obtained during increasing $\theta$, using a function composed of two antisymmetric Gaussian components, shown by the dashed and dotted lines.
}
\label{phib}
\end{figure}

To investigate the possible anomaly at $B_{\rm K}$ in $\214$, 
we measured the field-angle dependence of magnetostriction at 0.12~K and 1.4~T, 
where the sample is likely in the Abrikosov state at $\theta = 90^\circ$ for $B \parallel b$.
Figures~\ref{phib}(a) and \ref{phib}(b) present the magnetostriction data, $\Delta L_a(\theta)/L_a$, and its field-angle derivative, $\lambda_{\theta,a}=(\partial L_a/\partial \theta)/L_a$, respectively.
Tilting the magnetic field away from the $a$ axis toward the $c$ axis suppresses superconductivity around $\theta = 90^\circ \pm 1.5^\circ$.
The solid line in Fig.~\ref{phib}(b) shows a fit using the sum of two antisymmetric Gaussian functions, $g_1(\theta)+g_2(\theta)$,
where each $g_i(\theta)$ is defined as the difference between two Gaussian functions centered symmetrically at $\theta=90^\circ \pm \theta_i$.
Here, we obtain $\theta_1=1.05^\circ$ and $\theta_2=0.62^\circ$.
These components are shown as dotted and dashed lines, respectively, indicating the presence of two distinct anomalies associated with $\Hc2$ and $B_{\rm K}$.
This result suggests that the anomaly at $B_{\rm K}$ vanishes when the in-plane magnetic field is rotated by approximately $1^\circ$ toward the $c$ axis.

\begin{figure}[t]
\includegraphics[width=3.2in]{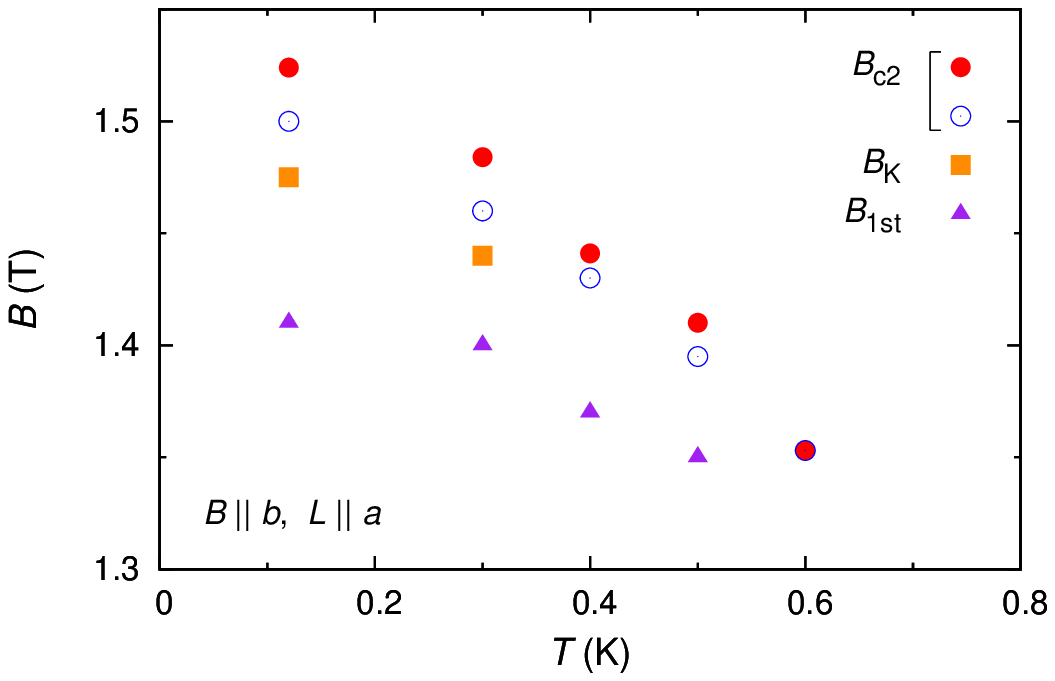} 
\caption{
(Color online) Field--temperature phase diagram determined from magnetostriction measurements for $L \parallel a$.
The closed (open) circles represent $\Hc2$ obtained during increasing (decreasing) magnetic field sweeps.
The squares indicate $B_{\rm K}$, where a hump-like anomaly is observed during increasing field.
The triangles mark $B_{\rm 1st}$, the field at which the hysteresis in $\Delta L_a^{\rm sc}(B)$ disappears.
}
\label{HT}
\end{figure}

\subsection{Thermodynamic relation associated with the magnetostriction jump}
For a first-order phase transition, the strain dependence of $\Hc2$ for $B \parallel a$, denoted as $B_{\rm c2}^{\parallel a}$, is governed by the Clausius-Clapeyron relation:
$\partial\Hc2^{\parallel a}/\partial\sigma_{xx}\approx\Delta \varepsilon_{xx}/\Delta M_{\parallel a}$,
where $\Delta \varepsilon_{ii}$ and $\Delta M_{\parallel a}$ correspond to the discontinuities across $\Hc2^{\parallel a}$ in $\Delta L_i/L_i$ and the $a$-axis component of the magnetization, respectively.
Using the experimentally observed values of $\Delta L_a/L_a \approx -2 \times 10^{-8}$ and $4\pi \Delta M_{\parallel a} \approx 0.7$~G,\cite{Kittaka2014PRB} we estimate 
$\partial\Hc2^{\parallel a}/\partial\sigma_{xx}\approx-0.3$~T/GPa in the zero-strain limit.
This estimate is of the same order as the value $(\partial\Hc2^{\parallel a}/\partial\sigma_{xx})|_{\sigma\rightarrow 0} \approx -0.1$~T/GPa, 
which is inferred under the assumption of a scaling relation 
$(\partial\Hc2^{\parallel a}/\partial\sigma_{xx})|_{\sigma\rightarrow 0} \sim (\partial\Tc/\partial\sigma_{xx})|_{\sigma\rightarrow 0} (\Hc2^{\parallel a}/\Tc)|_{\sigma\rightarrow 0}$, 
using the experimental observation of $(\partial\Tc/\partial\sigma_{xx})|_{\sigma\rightarrow 0}\approx -0.1$~K/GPa.\cite{Barber2019PRB}

\subsection{Possible origins of $B_{\rm K}$ anomaly}
The high-resolution magnetostriction measurements reveal a possible anomaly at $B_{\rm K}$ in $\lambda_a(B)$, slightly below the Pauli-limited upper critical field $\Hc2$ in $\214$. 
This feature, along with the double-peak structure observed in $\lambda_{\theta,a}$, 
suggests a lattice response that may be linked to the emergence of the FFLO phase. 
Notably, the anomaly at $B_{\rm K}$, situated within the hysteresis region above $B_{\rm 1st}$, is reminiscent of 
FFLO signatures reported in CeCoIn$_5$, 
which have been attributed to changes in the spatial modulation of the superconducting order parameter.\cite{Kittaka2023PRB}
In CeCoIn$_5$, the FFLO phase boundary is interpreted to correspond to the lower onset field of the hump-like anomaly.

However, a significant discrepancy exists between the FFLO phase boundary inferred from the magnetostriction data and that identified in recent NMR studies. 
While the NMR results indicate a FFLO boundary with a positive slope in the $B$--$T$ phase diagram,\cite{Kinjo2022Science} 
the magnetostriction measurements reveal negative slopes in both the $B_{\rm 1st}(T)$ and $B_{\rm K}(T)$ lines, as shown in Fig.~\ref{HT}. 
This inconsistency apparently underscores the probe-dependent nature of the  FFLO phase manifestation.
Moreover, the absence of a clear thermodynamic signature of the FFLO transition in specific-heat and entropy measurements implies the subtle nature of this phase. 

An alternative interpretation for the origin of the $B_{\rm K}$ anomaly is the broadening of the first-order transition.
Indeed, the presence of eutectic boundaries involving Sr$_3$Ru$_2$O$_7$ may influence the nucleation and stability of domains in which normal and superconducting states coexist. 
Although the FFLO state remains an intriguing possibility in $\214$, its realization has yet to be firmly established and warrants further investigation.

\section{Summary}
In this study, we performed high-resolution magnetostriction and thermal-expansion measurements on the Pauli-limited superconductor $\214$ using high-quality single crystals. 
Our results revealed a clear first-order superconducting transition under in-plane magnetic fields, accompanied by pronounced hysteresis 
and a subtle lattice response on the order of $10^{-8}$. 
A hump-like anomaly in the magnetostriction coefficient and a double-peak structure in its field-angle derivative were identified slightly below $\Hc2$, 
suggesting a possible link to the emergence of the FFLO phase. 
However, the observed features may also be interpreted as a broadening of the first-order transition.
Notably, no corresponding anomaly was detected in the magnetostriction measurements at the magnetic fields where NMR studies reported signatures of the FFLO phase.
This discrepancy underscores the need for further experimental and theoretical investigations to clarify its realization in $\214$.

\section*{Acknowledgments}
We thank A. P. Mackenzie, T. Terashima, K. Ishida, Y. Maeno, and Y. Shimizu for fruitful discussions.
In particular, we are grateful to A. P. Mackenzie for kindly supporting in the growth of high-quality single crystals in Dresden.
This work was supported by JST FOREST Program (JPMJFR246O), a Grant-in-Aid for Scientific Research on Innovative Areas ``J-Physics'' (JP15H05883, JP18H04306) from MEXT, Chuo University Grant for Special Research, 
and KAKENHI (JP17K05553, JP18K04715, JP20K20893, JP21K03455, JP23K25825, JP23H04868, JP23K22444, JP24K01461) from JSPS.

\clearpage
\onecolumn
\renewcommand{\thefigure}{S\arabic{figure}}
\setcounter{figure}{0}

\begin{center}
{\large Supplemental Material for \\
\textbf{High-resolution magnetostriction measurements of the Pauli-limited superconductor Sr$_2$RuO$_4$}}\\
\vspace{0.1in}
Shunichiro Kittaka,$^{1,2}$ Yohei  Kono$^2$, Toshiro Sakakibara,$^{3}$ Naoki Kikugawa,$^{4}$ Shinya Uji,$^{4}$ \\
Dmitry A. Sokolov,$^5$ and Kazushige Machida$^6$\\
{\small 
\textit{$^{1}$Department of Basic Science, The University of Tokyo, Meguro, Tokyo 153-8902, Japan}\\
\textit{$^{2}$Department of Physics, Faculty of Science and Engineering, Chuo University, Kasuga, Bunkyo-ku, Tokyo 112-8551, Japan}\\
\textit{$^3$Institute for Solid State Physics, The University of Tokyo, Kashiwa, Chiba 277-8581, Japan}\\
\textit{$^4$National Institute for Materials Science, 3-13 Sakura, Tsukuba, Ibaraki 305-0003, Japan}\\
\textit{$^5$Max Planck Institute for Chemical Physics of Solids, Nothnitzer Str. 40, 01187 Dresden, Germany}\\
\textit{$^6$Department of Physics, Ritsumeikan University, Kusatsu, Shiga 525-8577, Japan}\\
}
(Dated: \today)
\end{center}

\section*{I. Specific heat of the samples used in this study}
$\214$ is the $n=1$ member of the so-called Ruddlesden-Popper (RP) series ruthenates Sr$_{n+1}$Ru$_n$O$_{3n+1}$.
Large single crystals of these ruthenates can be grown by using a Ru self-flux floating-zone technique.
Due to evaporation of RuO$_2$ from the solvent at the melting point,
precise control of the growth parameters is essential for obtaining high-quality single crystals of $\214$~\cite{Mao2000MRBSM,Bobowski2019CMSM}.
As a result, Ru lamellae and/or $\327$ inclusions are frequently observed in single crystal rods of $\214$.
Figure \ref{CT}(a) shows the temperature dependence of $c_p/T$ at 0~T for three samples, labeled \#3-2, \#3-5, and \#3-B.
Sample \#3-2 was used in a previous specific-heat study \cite{Kittaka2018JPSJSM}, 
while samples \#3-B and \#3-5 were used for magnetostriction measurements in the present study.
In the main text, we present the results obtained using sample \#3-B.

As shown in Fig.~\ref{CT}(a), the Sommerfeld coefficient $\gamma_{\rm e}$, i.e., the normal-state value of $c_p/T$, 
is unexpectedly enhanced for sample \#3-B.
It is important to note that $c_p$ in Fig.~\ref{CT}(a) was evaluated using the following equation: 
\begin{equation}
c_p=\frac{c_{\rm raw}}{(m/M_{214})},
\end{equation}
where $c_{\rm raw}$ is the measured heat capacity in J K$^{-2}$, $m$ is the sample mass, and $M_{214}$ is the molar mass of $\214$.
If the $n=2$ member of the RP series, $\327$, is unintentionally included in the sample, 
the apparent normal-state $c_p/T$ is enhanced, since the Sommerfeld coefficient of $\327$ ($\gamma_{327} \sim 0.22$~J mol$^{-1}$ K$^{-2}$) is larger than that of $\214$ ($\gamma_{214} \sim 0.04$~J mol$^{-1}$ K$^{-2}$).
Indeed, polarized light optical microscopy images of the polished plane of sample \#3-B revealed
a small amount of $\327$ inclusions, as shown in the inset of Fig.~\ref{CT}(b).
To evaluate the specific heat of the $\214$ component in sample \#3-B, 
we define $c_{214}$ as:
\begin{equation}
c_{214}=\frac{c_{\rm raw}-\gamma_{327}T(m_{327}/M_{327})}{(m-m_{327})/M_{214}},
\end{equation}
where $m_{327}$ and $M_{327}$ denote the mass and molar mass of the $\327$ inclusions, respectively. 
We found that assuming $m_{327}=5.5$~mg (approximately 10\% of $m$) yields $c_{214}/T \sim 0.04$~J mol$^{-1}$ K$^{-2}$ in the normal state for sample \#3-B, as shown in Fig.~\ref{CT}(b).
For comparison, $m_{327}$ is assumed to be zero for samples \#3-2 and \#3-5.
Based on these analyses, we conclude that the enhancement of $c_p/T$ in sample \#3-B originates from $\327$ inclusions.
Nevertheless, the quality of the $\214$ component in sample \#3-B is comparable to that of samples \#3-2 and \#3-5,
as evidenced by the good agreement among the $c_{214}$ data for all three samples in Fig.~\ref{CT}(b).
To focus on the superconducting properties of the $\214$ component, we adopt $c_{214}$ in the main text.

\section*{II. Effect of $\Tc$ distribution on the magnetostriction}
To verify the reproducibility of the results, 
we performed magnetostriction measurements for $L \parallel c$  on sample \#3-5 in a dilution refrigerator, in addition to those on sample \#3-B shown in main text.
Sample \#3-5 ($m=12$~mg and $L_c=0.7$~mm) is smaller than sample \#3-B ($m=50$~mg and $L_c=1.1$~mm).
Although the onset $\Tc$ of sample \#3-5 is comparable to that of sample \#3-B, 
the specific heat jump at $\Tc$ is significantly sharper in sample \#3-5, as shown in Fig.~\ref{CT}(b);
the jump widths for samples \#3-5 and \#3-B are approximately 0.05 and 0.09~K, respectively.
These observations suggest that the $\Hc2$ distribution is considerably narrower in sample \#3-5.

Figures \ref{Mag}(a) and \ref{Mag}(b) show the relative capacitance change, $\Delta C=C(T,B)-C(T, 1.7\ {\rm T})$, for sample \#3-5,
measured at 0.3 and 0.09~K, respectively, under an in-plane magnetic field applied at $\phi=45^\circ$,
where the magnetic-torque effect is negligible.
The superconducting component of the relative magnetostriction, $\Delta L_c^{\rm sc}/L_c$, is plotted in Figs. \ref{Mag}(c) and \ref{Mag}(d) at 0.3 and 0.09~K, respectively.
The corresponding magnetostriction coefficient is shown in Figs.~\ref{Mag}(e) and \ref{Mag}(f).
Qualitatively similar behavior to that observed in sample \#3-B (see main text) was obtained.
A hump-like anomaly was not prominently detected in sample \#3-5 for $L \parallel c$, either.
After this measurement, sample \#3-5 was found to be cleaved,
suggesting the presence of internal cracks that may have contributed to signal scattering.
Nevertheless, these results demonstrate that the magnetostriction behavior for $L \parallel c$ is reproducible across different samples,
supporting the intrinsic nature of the observed features.

\clearpage

\setcounter{figure}{0}
\renewcommand{\thefigure}{S\arabic{figure}}
\begin{figure}[h]
\includegraphics[width=6.2in]{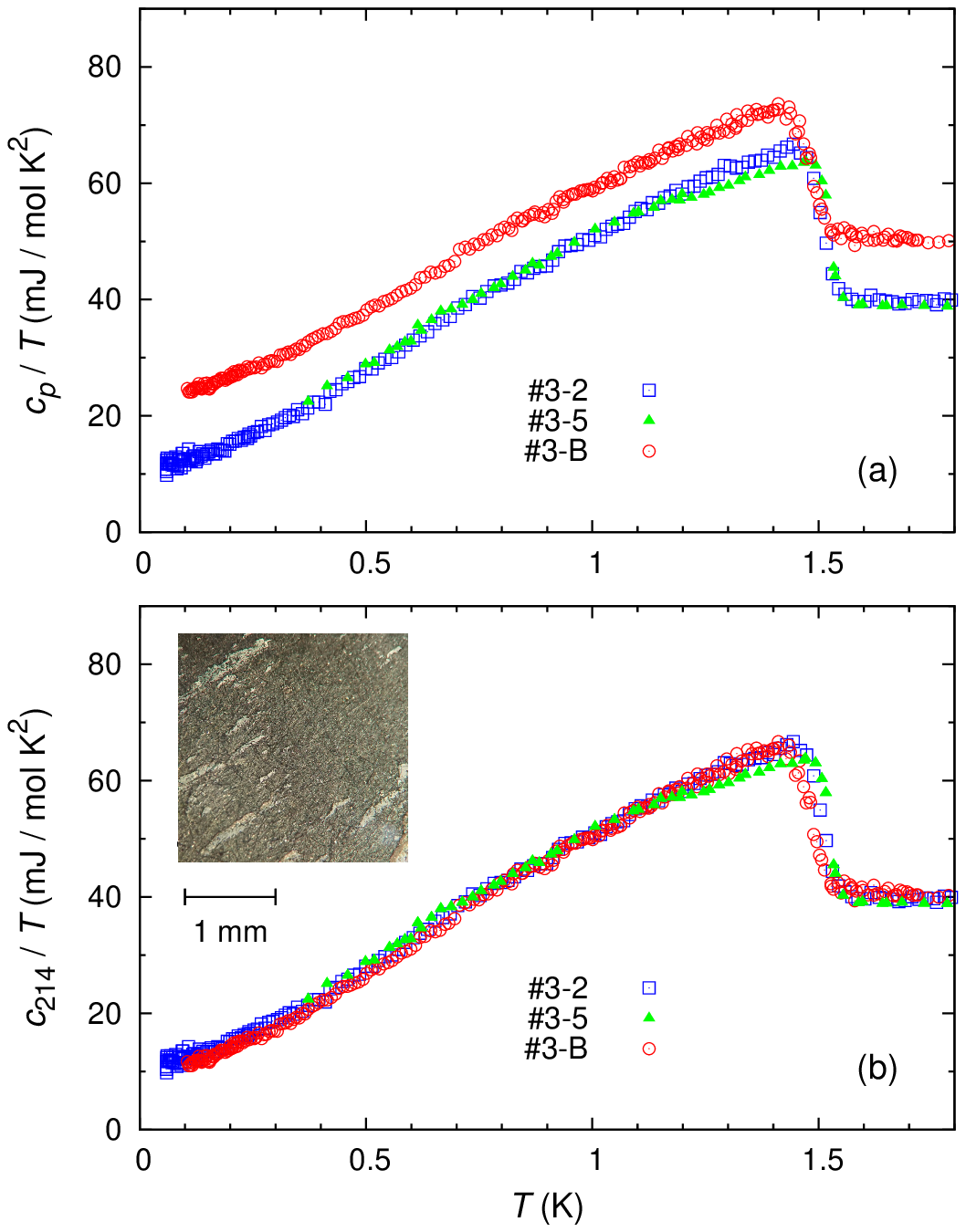} 
\caption{
Temperature dependence of the specific heat data, (a) $c_p/T$ and (b) $c_{214}/T$, in zero magnetic field for samples \#3-2, \#3-5, and \#3-B.
The inset in (b) shows a polarized light optical microscopy image of a polished surface of sample \#3-B.
The darker (brighter) area corresponds to $\214$ ($\327$).
}
\label{CT}
\end{figure}

\begin{figure}[h]
\includegraphics[width=6.2in]{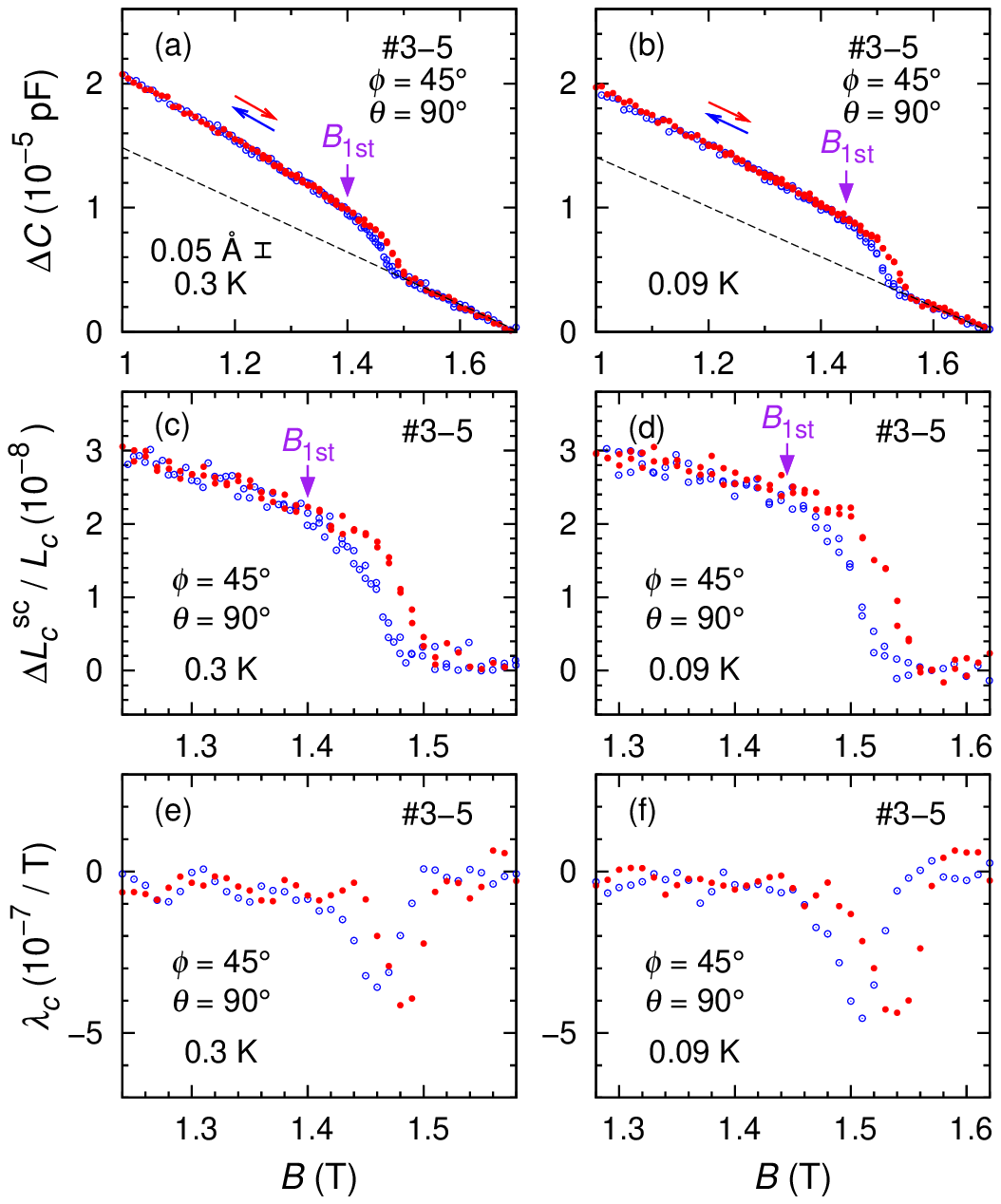} 
\caption{
Magnetic-field dependence of the capacitance change relative to the value at 1.7~T for sample \#3-5, 
measured at (a) 0.3 and (b) 0.09~K with $\phi=45^\circ$ and $\theta=90^\circ$,
during increasing (closed circles) and decreasing (open circles) field sweeps.
The dashed lines present linear fits to the data in the normal state, which are assumed to reflect the non-superconducting background contribution.
(c), (d) Superconducting component of the relative magnetostriction, $\Delta L_c^{\rm sc}/L_c$, obtained by subtracting the background contributions corresponding to the dashed lines in (a) and (b).
(e), (f) Magnetostriction coefficient $\lambda_c=(\partial \Delta L_c^{\rm sc}/\partial B)/L_c$ estimated from the data in (c) and (d).
}
\label{Mag}
\end{figure}


\begin{thebibliography}{99}
\bibitem{KittakaTorque}
The asymmetric behavior in $\Delta C(\theta)$ was also observed in another
  sample, which becomes negligible at $\phi=45^\circ$. The difference between
  the characteristic angles for these two samples ($\phi=27^\circ$ and
  $45^\circ$) evidences that the asymmetric behaivor comes from an extrinsic
  origin.
\end{thebibliography}
\end{document}